\baselineskip=20pt
\parindent 20pt
\settabs 4 \columns
\parskip=10pt 
\hfuzz=2pt   
\input epsf
\def\line{\hbox to \hsize}    
\def\frac #1#2{{#1\over #2}}

\def\Psid{\Psi^{\dagger}}
\def\psid{\psi^{\dagger}}


\line{\hfil May 1996}
\vskip 1cm
\centerline{\bf }
\centerline{\bf SPECTRAL FLOW, MAGNUS FORCE AND MUTUAL FRICTION VIA}
\centerline{\bf THE GEOMETRIC OPTICS LIMIT OF ANDREEV REFLECTION}
\vskip 1cm
\centerline{Michael Stone}
\vskip .5cm                                
\centerline{\it University of Illinois at Urbana Champaign}
\centerline{\it Department  of Physics}
\centerline{\it 1110 W. Green St.}
\centerline{\it Urbana, IL 61801}
\centerline{\it USA}
\vskip 1cm   

\line{\bf Abstract\hfil}

The notion of spectral flow has given new insight into the motion of
vortices in  superfluids and superconductors.  For a BCS
superconductor the  spectrum of low energy vortex core  states is
largely  determined by the geometric optics limit of Andreev
reflection. We use this to follow the evolution of the
states when a stationary vortex is immersed in a transport
supercurrent.  If the core spectrum were continuous, spectral flow
would  convert the momentum flowing into the core via the Magnus
effect into unbound quasiparticles --- thus allowing the vortex to
remain stationary without a pinning potential or other sink for the
inflowing momentum. The discrete nature of the states, however, leads
to Bloch oscillations  which thwart the spectral flow. The momentum
can escape only via relaxation processes.  Taking these into account
permits a physically transparent derivation of the mutual friction
coefficients.

\vfil\eject

\line{\bf 1) Introduction\hfil}

Imagine a two dimensional superfluid, initially in its ground state,
confined to the surface of a torus.  Suppose now that a
vortex-antivortex pair is created at some point on the surface and
the vortex is moved slowly round one of the generators of the torus
before being allowed to annihilate with its antivortex partner.  One
effect of this process is to give the superfluid order-parameter phase a
unit winding number around the generator perpendicular to the motion
of the vortex. The associated phase gradient implies that a
supercurrent has been established in this direction. If no other
momentum-carrying excitations were created along with the
supercurrent, the system as a whole has acquired momentum
perpendicular to the vortex motion.  Moving the vortex  therefore
requires us to supply this momentum from an external source.  This is
the Magnus effect [1].

For a Bose superfluid this is all there is to the story: If we wish
to move a vortex with respect to the background fluid  we must (at
least at low temperatures when there is no normal fluid component)
place a wire or other object in the vortex core to supply the
transverse momentum to the fluid. The reaction  force the fluid
exerts on the wire is the Magnus, or Kutta-Joukowski, lift force[2].

For a  fermionic $S$-wave superfluid\footnote {*}{We consider a
neutral condensate for simplicity.  The principal effect of the
magnetic field in an Abrikosov vortex is to transfer the momentum
supplied by the vortex to the positive ion lattice, thus ensuring
that no superflow is induced beyond the penetration depth.  Nothing
significant changes in the core.} the situation is subtler because
the vortex has low-energy bound states [3,4] whose role in
the momentum balance equation has been studied for many years
[5,6,7].  Recently Volovik [8] has cast a new light on this subject
by showing that motion of the vortex with respect to the stationary
condensate induces a spectral flow among these states.  In a cartoon
version of his theory this spectral flow  generates, even with
adiabatic motion of the vortex, a stream of unbound quasi-particles
which carry off momentum equal and opposite to that of the induced
superflow. The vortex can apparently be moved without any external
source of transverse momentum. In this sense the spectral flow
``cancels'' the Magnus effect.  The discrete nature of the bound
state spectrum, however, complicates the picture.  As observed in [8]
and modeled in [9], a non-zero temperature is required to broaden the
closely spaced levels so that they may behave as if the spectrum were
continuous. It is only in the hydrodynamic limit that 
``cancellation''
takes place.

Despite the complicating necessity of level broadening, the spectral
flow mechanism provides a very physical picture of the processes
occuring in the vortex core.  Several questions immediately arise,
however. For example: What happens when a vortex is held stationary
in a transport supercurrent?  By galilean invariance, this situation
is physically equivalent to a moving vortex and a stationary
superfluid, yet --- with no time dependence in the Bogoliubov-de
Gennes equations --- it is not immediately clear what drives the
spectral flow.  A second question is whether the spectral
flow picture requires  a modification of the conventional theories of
momentum balance and  Hall angle. The aim of the present paper is to
discuss these issues within a simple model for the core states.

We will introduce a quasi-classical picture, based on the geometric
optics limit of Andreev  scattering [10], for the evolution of the
states in a stationary vortex core. This allows us to show that the
discreteness of the spectrum leads to effects analogous to those in
the one-dimensional Wannier-Stark ladder.  Stark ladder resonances
occur when a uniform electric field is applied to a Bloch electron in
a periodic potential [11].  The electron initially accelerates but,
in the absence of dissipation, is eventually Bragg reflected from the
periodic lattice potential resulting in an oscillatory motion in a
localized region.  No net current flows --- except that arising from
the exponentially small inter-band Zener tunneling. If we describe
this process in a gauge where $A_0=0$, $A_x=-Et$ we have a
time-dependent hamiltonian and explicit spectral flow.  In a gauge
where $A_0=-Ex$, $A_x=0$ there is no explicit time dependence but the
same physics results.  In both gauges dissipation via inelastic
collisions allows the electron to avoid Bragg reflection and so
permits a finite current.  Analogously, relaxation processes in the
vortex core allows some spectral evolution. By keeping track of the
resultant momentum flux we find the consequences of the spectral flow
for the vortex dynamics.  These turn out to be  the well-known mutual
friction that couples the superflow to the normal flow via the vortex
motion.  The traditional Green function formalism of refs [6,7] must
therefore tacitly take the spectral flow into account.

The organization of this paper is as follows:  First, in section 2),
we exhibit a simple version of spectral flow and show how momentum
entering the vortex core is recycled as quasi-particle momentum. In
section 3) we will review  the theory of the core states and its
connection with Andreev reflection.  In 4) we interpret the core
state spectrum in terms of the failure of exact Andreev
retro-reflection and, in 5), armed with the insight gained from this
interpretation, we show how the spectral flow  is mapped onto the
Stark-Wannier problem. Finally,  also in section 5), we account for the 
momentum flux  to the normal component.

\vskip 12pt

\line{\bf 2) One-dimensional Spectral Flow\hfil}

As a simple model of a vortex core 
consider the 
one-dimensional Dirac-Andreev eigenvalue problem
$$
\left[\matrix {-iv_f\partial_x & \Delta(x)e^{i\theta(x)}\cr
                    \Delta(x) e^{-i\theta(x)} & iv_f\partial_x\cr}\right]
\left[\matrix{ u\cr v\cr}\right]
=\epsilon\left[\matrix{u\cr v\cr}\right],
\eqno(2.1)
$$
where $\Delta(x)=0$ for $0<x<L$ (the ``core'') while $\Delta(x)=\Delta=const.$ elsewhere. 
We take the phase of the order parameter to be $\theta(x)=\theta_L$
for $x <0$
and $\theta(x)=\theta_R$ for $x>L$. Here $v_f$ is the Fermi velocity. 
We will use $m$ to denote the fermion mass so that $k_f= mv_f$ is the Fermi
momentum and $E_f=\frac 12 mv_f^2$ is the Fermi energy.

The bound-state solutions, $\Psi= \left[\matrix{ u\cr v\cr}\right]$, 
with $\epsilon<\Delta$ are easily found. The
wavefunctions are of the form
$$
\eqalign{
\Psi(x)=&\left[\matrix{\epsilon+ik v_f\cr \Delta e^{-i\theta_R}\cr}\right] e^{-k(x-L)} 
\qquad x>L,\cr
=&\left[\matrix{a e^{i\epsilon x/v_f}\cr b e^{-i\epsilon
x/v_f}\cr}\right]
\quad\qquad \qquad 0<x<L.\cr
=&\left[\matrix{\epsilon-ik v_f\cr \Delta e^{-i\theta_L}\cr}\right] e^{kx} 
\qquad \qquad x<0,\cr
}
\eqno(2.2)
$$
with $\epsilon^2+(v_f k)^2=\Delta^2$.
Matching the solutions at $x=0,L$ fixes the ratio $a/b$ and requires the
eigenvalue $\epsilon_n$ to obey
$$
\epsilon_n = \frac {v_f}{2L}\left( (\theta_R-\theta_L)+2\pi n
+2\cos^{-1}(\epsilon_n/\Delta)\right)
\eqno(2.3)
$$
For states deep in the gap, $\epsilon\ll \Delta$, this simplifies to
$$
\epsilon_n = \frac {v_f}{2L}\left( (\theta_R-\theta_L)+2\pi (n+\frac 12)\right)
\eqno(2.4)
$$

We see that if we gradually increase the phase difference across the
core, $\Delta\theta=\theta_R-\theta_L$, the entire spectrum moves up
in energy. By the time $\Delta\theta$   has increased by $2\pi$  each
state has been replaced by the one below it. This is the spectral flow.

The physical interpretation of the spectral flow depends on the context. If (2.1) were
describing a charge density wave (CDW) system, the upper and lower components
of $\Psi$ would be the amplitude of left- and right-going particles.  
Then a summation over occupied states gives the local charge density and current
$$
\eqalign{
<\Psid (x)\Psi (x)>=&<\psid_R(x)\psi_R(x)+\psid_L(x)\psi_L(x)>=<\rho(x)>\cr
v_f<\Psid (x) \sigma_3\Psi(x)>
= &v_f<(\psid_R(x)\psi_R(x)-\psid_L(x)\psi_L(x))>=<j(x)>.\cr
}
\eqno(2.5)
$$

In a CDW a time rate
of change
of the phase of the order parameter induces a current $<j>\approx \frac
1{2\pi} \dot\theta $, with the corrections being small when $\dot\theta$ is small. 
Consequently the slow twisting of $\theta_R$ relative
to $\theta_L$ tells us that charge is flowing into the the gapless 
region $0<x<L$.
Since the time-dependent version of the Dirac-Andreev equation implies that
$\rho$ and $j$ obey the conservation law
$$
\partial_t \rho (x)+\partial_x j(x)=0,
\eqno(2.6) 
$$
the inflowing charge must be accumulating in the gapless region
[12].
Each time the relative twist increases by $2\pi$, a unit charge will
have accumulated. In the same interval 
one  of the (occupied) negative energy bound state levels has
adiabatically crossed the zero energy level and taken the place
of a positive energy state. The occupation number of the
positive energy bound states has therefore increased by unity,
consistent with the accumulation of unit charge.
Eventually, with more twisting (the amount depending on $L$),
the filled levels will reach the top
of the gap and merge with the upper continuum. After this point each
new unit of charge that flows in will appear as a low energy
quasi-particle.

In a superconductor the upper and lower components of $\Psi$ are
$\psi_R$, $\psid_L$ respectively. In this case  the 
expressions for the current and charge density are interchanged
$$
\eqalign{
v_f<\Psid (x)\Psi(x)>
=&v_f<\psid_R(x)\psi_R(x)+\psi_L(x)\psid_L(x)>\cr
=&v_f<\psid_R(x)\psi_R(x)-\psid_L(x)\psi_L(x)>=<j(x)>\cr &\cr
\Psid (x) \sigma_3\Psi(x)
= &<\psid_R(x)\psi_R(x)-\psi_L(x)\psid_L(x)>\cr
= &<\psid_R(x)\psi_R(x)+\psid_L(x)\psi_L(x)>\cr=&<\rho(x)>.\cr
}
\eqno(2.7)
$$
\noindent 
The conservation law
$$
\eqalign{
&\partial_t(\psid_R\psi_R-\psid_L\psi_L) +v_f\partial_x
(\psid_R\psi_R+\psid_L\psi_L)\cr
&=\partial_t\Psid\Psi +v_f\partial\Psi\sigma_3\Psi\cr &=0,\cr}
\eqno(2.8)
$$
\noindent also changes its physical interpretation as, on
multiplication by $k_f$, it becomes the equation of momentum
conservation [13]. Now, instead of charge, each occupied bound state
carries momentum $+k_f$.  The relative twisting of the phases on the
two sides of the core represents an inflow of momentum from the
condensate, and the spectral flow leads to its recycling as the
momentum of low energy quasi-particles.

We can make a simplistic model of the consequences of two-dimensional
vortex motion by assuming that (to a first approximation) the process
described in the introduction, the passage of a vortex around the
$L_y$ generator of an $L_x\times L_y$ torus, can be mimicked as the
breaking and reconnection after a $2\pi$ phase twist of the order
parameter in a collection of one-dimensional superfluids, one for
each allowed $k_y=2\pi n/L_y$. From the discussion following (2.8) we
see that a single twist accumulates a momentum $k_x=\sqrt{|k_f|^2
-k_y^2}$ for each  of the one-dimensional systems. This is the
same amount of momentum we would get by translating the  $k_x$ value
of each particle on the torus by $\delta k_x=\pi/L_x$. If the total
number of electrons is $N$, the net momentum accumulated in a passage
of a vortex round the $L_y$ generator is therefore
$$
\Delta P_x= \frac{\pi N}{L_x}= \frac {\pi \rho}{m} L_y,
\eqno (2.9)
$$
where $\rho=m N/(L_x L_y)$ is the mass density.

This is implies a rate of momentum accumulation in the core of
$$
\frac {dP_x}{dt}= \frac {\pi}{m} \rho v_y.
$$
Since the circulation in a BCS vortex is $\kappa=\pi/m$ we see that
momentum is accumulating, or being recycled, at a rate equal to the Magnus
force on the vortex, $F_x=\kappa \rho v_y$.

This cartoon version of the process is of course overly simplistic. 
The bound states in the
two dimensional vortex core are not those of the one-dimensional
equation (although we will soon see that they are closely related),
and the variation of the order parameter in the $y$ direction
will couple the $k_y$ momenta so that they cannot be dealt with
individually. In the next section we will begin to deal with these
deficiencies.

\vskip 12pt

\line{\bf 3) Two dimensional bound states\hfil}

In this section we will review the classical results of [3,4] on bound
states in the   core of a two-dimensional vortex. Our aim is to show that
the physics of Andreev scattering  
reduces the full problem to a collection of one dimensional
problems of the form considered in the previous section. 

We wish to solve the Bogoliubov-de Gennes equation
$$
\left [\matrix{-\frac  1{2m}\nabla^2 -E_f & \Delta(r) e^{i\theta}\cr
               \Delta(r) e^{-i\theta} & \frac  1{2m}\nabla^2 +E_f\cr}\right]
\left[\matrix{\tilde u \cr \tilde v\cr}\right] =\epsilon \left[\matrix{\tilde  u \cr
\tilde v\cr}\right].
\eqno (3.1)
$$
Here $r$ and $\theta$  are polar coordinates with origin at the vortex
centre.

For the moment we will leave the gap profile $\Delta(r)$ unspecified,
but the angular dependence of the order parameter 
is such that the superflow is {\it anti-clockwise} with a single quantum of circulation. 

We now separate the radial and angular parts of the wavefunction. To
do this we must first appreciate that $u$, $v$ are invariant only under
$4\pi$ rotations\footnote{*}{The same is true for the one dimensional
problem in section 2). There we saw that a $2\pi$ twist in the order
parameter shifts the particle momenta by $\delta k_x =\pi/L_x$. If
there were no quasi-particle created along with the twist this would
lead to a double valued many-body wavefunction. Fortunately the
$x$ dependence of the quasi-particle serves to restore the single valuedness
to the total wavefuction.}. Therefore we  seek
solutions in the form
$$
\left[\matrix{\tilde  u \cr \tilde  v\cr}\right]=
\left[\matrix{ u(r,\theta) e^{i\theta/2} \cr v(r,\theta) e^{-i\theta/2}\cr}\right]=
\left[\matrix{ u_l(r) e^{i\theta/2} \cr v_l(r) e^{-i\theta/2}\cr}\right]
e^{i l\theta} \qquad l \in {\bf Z}.
\eqno (3.2)
$$
We find that $u_l(r)$, $v_l(r)$ obey
$$ 
\left [\matrix{-\frac  1{2m}(\partial_{rr}^2+\frac 1r \partial_r-
\frac {(l+\frac 12)^2}{r^2}+k_f^2) & \Delta\cr
\Delta &\frac 1{2m}(\partial_{rr}^2+\frac 1r \partial_r-\frac {(l-\frac 12)^2}{r^2}+k_f^2) }\right]
\left[\matrix{u_l \cr v_l\cr}\right] =\epsilon \left[\matrix{ u_l \cr
vl\cr}\right].
\eqno (3.3)
$$

To separate the rapidly varying degrees of freedom
from the slow, we follow [3] and  write
$$
\left[\matrix{ u_l \cr
v_l\cr}\right]= \left[\matrix{ f_l \cr
g_l\cr}\right]H_m^{(1)}(k_fr) +  c.c
\eqno (3.4)
$$
where $H_m^{(1)}(k_fr)$ with $m=\sqrt{ l^2+\frac 14}$ is a Hankel function,
and $f_l$, $g_l$ are slowly varying envelope functions.
We may approximate the Hankel function by the WKB form
$$
H_m^{(1)}(k_fr)\sim \frac {const.}{(r^2-b^2)^{\frac 14}} \exp \left\{
ik_f\int_b^r \frac {\sqrt{r^2-b^2}}{r} dr\right\},
\eqno (3.5)
$$
which is valid for $r$ larger than the WKB turning point
$r=b$. This turning point is related to 
$k_f$ and $l$ by $k_fb=l$. 

Substituting  (3.5) in (3.3) and keeping only the large terms we find
$$
\left [\matrix{ -iv_f \frac {\sqrt{r^2-b^2}}{r} \partial_r & \Delta\cr
\Delta & iv_f \frac {\sqrt{r^2-b^2}}{r} \partial_r }\right]
\left[\matrix{f_l \cr g_l\cr}\right] =
\left(\epsilon- \frac l{2m r^2 }\right) \left[\matrix{ f_l \cr
g_l\cr}\right].
\eqno (3.6)
$$

We can make (3.6) more intelligible by trading 
$r$ for a variable $x$ defined by $r^2=x^2+b^2$ [4]. We find
$$  
\left [\matrix{ -iv_f  \partial_x & \Delta\cr
\Delta & iv_f \partial_x }\right]
\left[\matrix{f \cr g\cr}\right] =
\left(\epsilon- \frac l{2m (x^2+b^2) }\right) \left[\matrix{ f \cr
g\cr}\right].
\eqno (3.7)
$$

This  substitution has a geometric meaning: For any Bessel function
$J_l(kr)$ the turning point distance, $b$, is the   impact parameter
of a particle moving  past the origin in a straight line with
momentum $k$  and angular momentum $l$.  The quantity $x$ is the
distance of the particle from the point of closest approach to the
origin.  Along with $x$ we introduce a polar angle ``$\theta$''  by
$x=b\tan \theta$.  This angle differs only by a constant from the
true polar angle $\theta$, so we temporarily  abuse notation and make
no distinction between the new angle and the original. We then notice
that
$$
\frac l{2m (x^2+b^2) }= \frac 12 v_f \frac {d\theta}{dx}
\eqno (3.8)
$$
so that, defining,
$$
\left[\matrix{ \tilde f_l \cr
\tilde g_l\cr}\right]= \left[\matrix{  f_l e^{i\theta/2} \cr
 g_l e^{-i\theta/2} \cr}\right]
\eqno (3.9)
$$
to undo the transformation that  removed the angle dependence  
from the order parameter in (3.1), we  get
$$ 
 \left [\matrix{ -iv_f  \partial_x & \Delta e^{i\theta}\cr
\Delta e^{-i\theta} & iv_f \partial_x }\right]
\left[\matrix{\tilde f_l \cr \tilde g_l\cr}\right] =\epsilon \left[\matrix{\tilde  f_l \cr
\tilde g_l\cr}\right].
\eqno (3.10)
$$

This looks very much like the one dimensional eigenvalue problem solved in
section 2). It is not yet identical, however. In (3.10) the coordinate
$x=\sqrt{r^2-b^2}$ is restricted to positive values. Furthermore  boundary
conditions have to be imposed on $\tilde f_l$, $\tilde g_l$ at $x=0$
to ensure that the Hankel functions can combine to give  the $J_{l\pm
\frac 12}(kr)=\frac 12(H_{l\pm
\frac 12}^{(1)}(kr)+H_{l\pm
\frac 12}^{(2)}(kr))$ which are
finite at the origin. We may nonetheless extend $x$ to negative values
by regarding the part of (3.4) with the incoming Hankel function
$H_m^{(2)}(k_fr)$ as living on the negative $x$ axis, and the
outgoing part $H_m^{(1)} (k_fr)$ on the positive $x$ axis.  With this interpretation
the boundary conditions at $x=0$ translate into the requirement of continuity of
$\tilde f_l$, $\tilde g_l$ there.

Physically the transformation of the two-dimensional eigenvalue
problem (3.1) into the one-dimensional (3.10) occurs because each
bound quasi-particle is bouncing back and forth along a straight line,
its direction of motion being repeatedly reversed 
by Andreev scattering off the increasing value of the gap.

\vskip 12 pt

\centerline{\epsfxsize=3.4in\epsffile{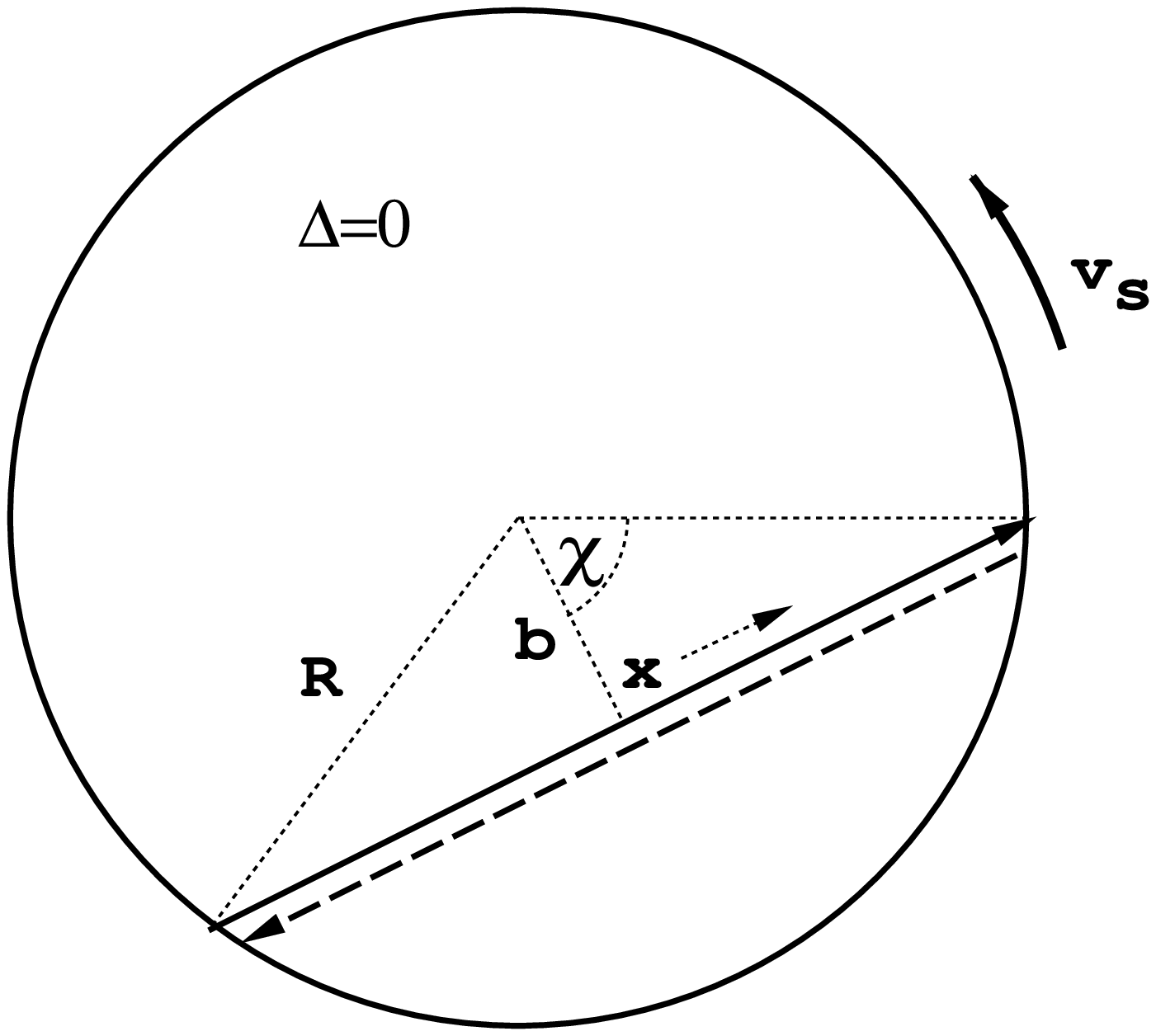}}

{\narrower\smallskip\noindent Fig. 1. A bound state with an electron (solid
arrow) being Andreev reflected as a hole (dashed arrow).\smallskip}

The $\Delta(r)$ profile found  from a self-consistent solution of the
Bogoliubov-de Gennes and gap equations has scale $R\approx v_f/\Delta$
[14].  In order to have analytic expressions for the bound state
eigenvalues, we will not use such a self consistent $\Delta(r)$, but
take 
instead a step function: $\Delta(r)=0$,
$r<R$ and $\Delta(r)=\Delta$ for $r>R$. We will also assume $R$ to be
somewhat larger than $v_f/\Delta$ so that we can ignore, in (3.10),
the variation of $\theta$ in the regions where the wavefunction is
evanescent. With these assumptions we can directly apply (2.3),
(2.4).

Each quasi-particle trajectory now coincides with a chord of a circle of radius
$R$.  If this chord subtends an angle of $2\chi$ at the centre of the
circle it has length $L=2R\sin\chi$ (see Fig. 1.). Furthermore the quantity
$\Delta \theta=\theta_L-\theta_R$, the difference in order parameter
phase at the two ends, is equal to $2\chi$.  For states with energy
$\epsilon \ll \Delta$ the energy eigenvalues are therefore 
$$
\epsilon_n(l)= \frac {v_f}{4R\sin\chi} (2\chi + 2\pi(n+\frac 12)).
\eqno(3.11) 
$$ 
Here $l=b k_f= R k_f\cos \chi$. For small $l$ and
$n=-1$, (3.11)  reduces to 
$$
\epsilon_{-1}(l)=-\omega_0 l, \qquad \omega_0= \frac {1}{2mR^2}.
\eqno (3.12) 
$$
This is the topological branch of low energy excitations which is
important for the spectral flow.  Note that $\omega_0=(2mR^2)^{-1}$
is the angular velocity of the superflow at the boundary  of the core.

The maximum value of $l$ occurs where the chord length becomes zero
and $\epsilon \to \Delta$. This maximum value is $l_{{\rm max}}= R
k_f \approx v_f k_f/\Delta = 2 E_f /\Delta \approx 10^3$---$10^4$.
This is large enough that $l$ can almost be regarded as a continuous
parameter. We will often use this observation to write expressions such as 
$\partial \epsilon/\partial l$ without further comment. 

In the next section we will show that the $l$ dependence in (3.11),
(3.12) is a consequence of the failure of the Andreev scattering to be
perfectly retro-reflective [15].

\vskip 12pt

\line{\bf 4) Andreev Reflection and the Bound State Spectrum\hfil}

The  bound states may be thought of as  standing waves set up by 
Andreev reflected quasi-particles repeatedly
traversing a chord of a circle. 
For a bound state of definite
angular momentum, $l$, the  orientation of this chord will be
indefinite. To produce a state localised on a chord
with specified orientation and impact parameter 
we must take a linear combination of angular momentum states,
$$
|\theta>=\sum_l a_l e^{-il\theta}|l>,
\eqno (4.1)
$$ 
where the coeficients $a_l$ are large only for $l\approx b/k_f$.
Because of  the Rayleigh criterion relating the extent of a plane
wave-front and its diffraction spread, there will be an uncertainty
relation between the sharpness of $\theta$ and $b$.

We may follow the time evolution of these wave packets by using the
small $l$ approximation $\epsilon_{-1}(l)=-\omega_0 l$.
We find
$$
|\theta,t>= \sum_l a_l e^{-il\theta}e^{+ il\omega_0  t}|l>=|(\theta-\omega_0
t),t=0>.
\eqno (4.2)
$$
The chord is thus seen to precess at the angular frequency
$\omega_0$ in the sense {\it opposite} to the superflow. If we use
the more general form (3.11) for the energies, we will need to 
replace $-\omega_0$ by a group angular velocity 
$$
\omega=\frac{\partial \epsilon(l)}{\partial l},
\eqno (4.3)
$$
but the same general picture will hold. In this section we will see
that this precession can be understood rather precisely as the
combined effect of two processes each of which causes  the Andreev reflection
to fail to be be perfectly retro-reflective. One of these processes
is due to the superflow and we will be able to use this insight to
determine the evolution of  the bound states when the vortex is
immersed in a transport supercurrent.

To begin with we consider a plane SNS junction of width $L$ having a
supercurrent $v_s$ flowing parallel to the junction (see Fig. 2.).

\vskip 12pt

\centerline{\epsfxsize=3.4in\epsffile{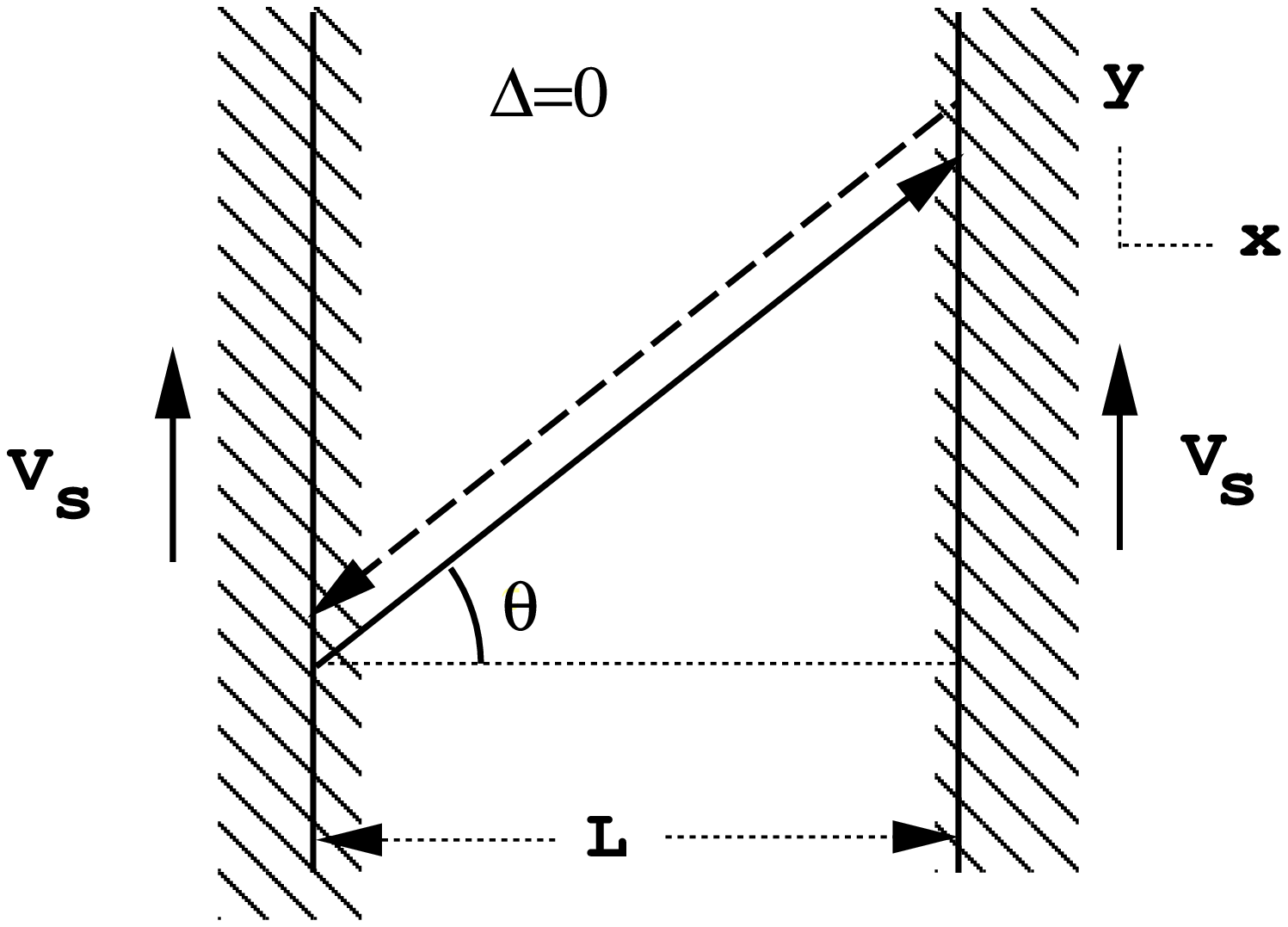}}

{\narrower \smallskip\noindent Fig. 2. Bound state in an SNS Junction with superflow. The
solid arrow is the electron trajectory and the dashed arrow the
hole.\smallskip}

There will be bound states in the junction with momenta close to
${\bf k}=(k_x,k_y)=(k_f\cos \theta, k_f\sin\theta)$.
These states are found from the Andreev hamiltonian
$$
H=\left [\matrix{ -i( k_x\partial_x+k_y\partial_y) & \Delta(x) e^{2imv_sy}\cr
\Delta(x) e^{-2imv_sy} & i( k_x\partial_x+k_y\partial_y) }\right],
\eqno (4.4)
$$
where $\Delta(x)=0$ for $0<x<L$ and $\Delta(x)=\Delta$ elsewhere.
They have the form
$$
\psi=\left[\matrix{ e^{i(\delta k_x x +\delta k_y y)} \cr
 e^{-i(\delta k_x x +\delta k_y  y)}\cr}\right], \qquad 0<x<L.
\eqno (4.5)
$$
Here $\delta k_y= mv_s$ and (for states with $\epsilon \ll \Delta$) $\delta k_x=\frac
1{2L}[2\pi(n+\frac 12)]$.
Their energy is 
$$
\epsilon= v_f( \delta k_x  \cos\theta + \delta k_y\sin \theta).
\eqno (4.6)
$$

In the figure we have drawn the trajectories of the electron and hole
as if they were at a definite location in the junction. In reality,
specifying a $y$ location requires making a wave packet 
$$
|y>=\int \frac {dk_y}{2\pi} a(k)e^{-ik_yy}|k>,
\eqno (4.7)
$$
with some spread of $k_y$.
This packet will drift in the $+y$ direction at the group velocity
$$
v_{\rm{drift}}=\frac{\partial \epsilon(k)}{\partial k_y} =-\frac 
{\tan\theta \delta k_x}{m} + v_s.
\eqno (4.8)
$$
For trajectories that are at close to normal incidence on the
supercurrents ($\theta=0$ or $k_y\approx 0$) this velocity is essentially $v_s$.

\vskip 12 pt

\centerline{\epsfxsize=3.4in\epsffile{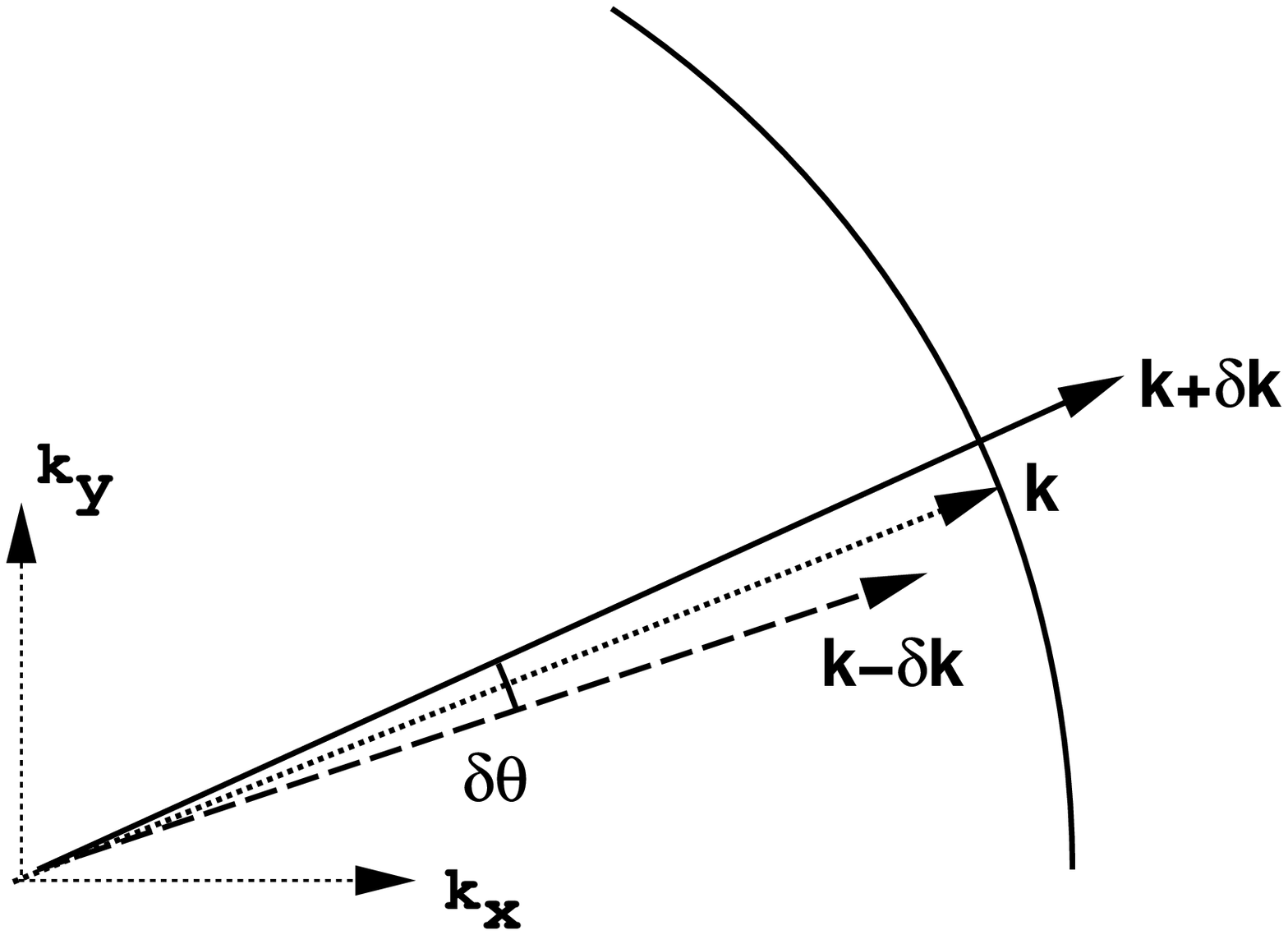}}

{\narrower\smallskip\noindent Fig. 3. The electron (solid arrow) and hole (dashed arrow)
momenta do not lie exactly at ${\bf k}$ on the Fermi surface, consequently their 
momenta differ in direction by a small angle $\delta\theta$.\smallskip}

This drift can be accounted for by noting that the Andreev
scattering is not perfectly retro-reflective. Consider reflection  
from the right-hand supercurrent in Fig. 2.  Although both the
incident electron and the reflected hole have nominal momentum ${\bf
k}$,  they in fact have momenta ${\bf k}\pm \delta {\bf k}$ (see Fig.3.). 

This
means that each reflection causes a small change in the angle of incidence 
$$
\eqalign{
\delta \theta = & 2(-\delta k_x \sin\theta +\delta k_y\cos\theta)\cr
=& - \frac {\epsilon}{E_f} \tan\theta + \frac {2
v_s}{v_f}\frac 1 {\cos \theta}.\cr
}
\eqno (4.9)
$$
Because both particle and hole move with speed $v_f$, 
it is easy to see (see Fig.3) that such a change in angle leads to the trajectory
migrating up the junction with velocity
$$
v_{\rm{drift}}=\frac {v_f }{2\cos\theta}\delta\theta.
\eqno (4.10)
$$
Inserting (4.9) in (4.10) precisely reproduces (4.8).

\vskip 12 pt

\centerline{\epsfxsize=3.4in\epsffile{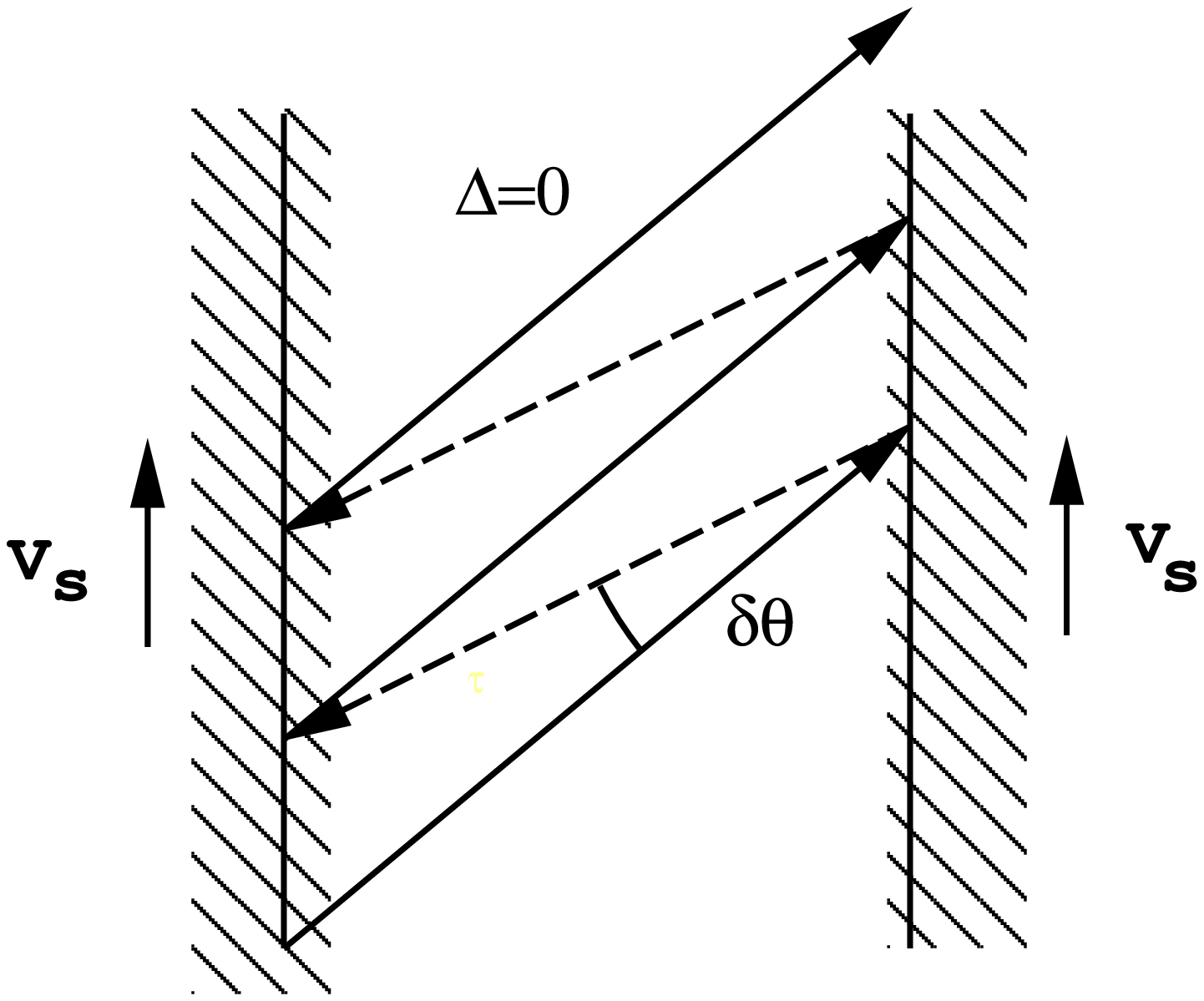}}

{\narrower\smallskip\noindent Fig. 4. The slight difference between the angle of
incidence of the electron (solid arrow) and the angle of reflection of the hole
(dashed arrow) causes the bound state to migrate up the SNS
junction.\smallskip}

The same result holds true for the bound states in the circular core.
From section 3) we know that the bound state energies are
$$
\epsilon_n(l)=\frac {v_f}{4 R \sin \chi}[(2\chi+ 2\pi(n+\frac 12)]
\eqno (4.11)
$$
where $l=k_f R\cos \chi$.
The group angular velocity is therefore
$$
\omega_g=\frac{\partial \epsilon}{\partial l}=\frac {\cos \chi}{\sin^2 \chi}\frac
{\epsilon}{k_f R} -\frac 1{2mR^2 \sin^2 \chi}.
\eqno (4.12)
$$
For states with small $\epsilon$ ({\it i.e.\/} $n=-1$, $\chi\approx \pi/2$)
the latter term dominates. 

The geometric optics picture works here also. We can use the plane
interface formula for  $\delta \theta$  after noting that
  the angle of incidence, called $\theta$
in (4.9) is now $\theta=\frac {\pi}{2}- \chi$. 
Notice, though,  that a positive  $\delta \theta$  means that the point
of impact of the return trajectory moves {\it backwards} through an
angle of $2\delta \theta$  (See Fig.5.). This is the origin of the minus
 sign in $\epsilon(l)\approx -\omega_0 l$. 
The other infomation we need is the time between impacts at the 
points $A$ and $C$. This is
 $\delta t=4 R \sin \chi/v_f $.
Putting all the ingredients together gives
$$
\omega=\frac {\cos \chi}{\sin^2 \chi}\frac
{\epsilon}{k_f R} -\frac 1{2mR^2 \sin^2 \chi}
\eqno (4.13)
$$ as before.

\centerline{\epsfxsize=3.4in\epsffile{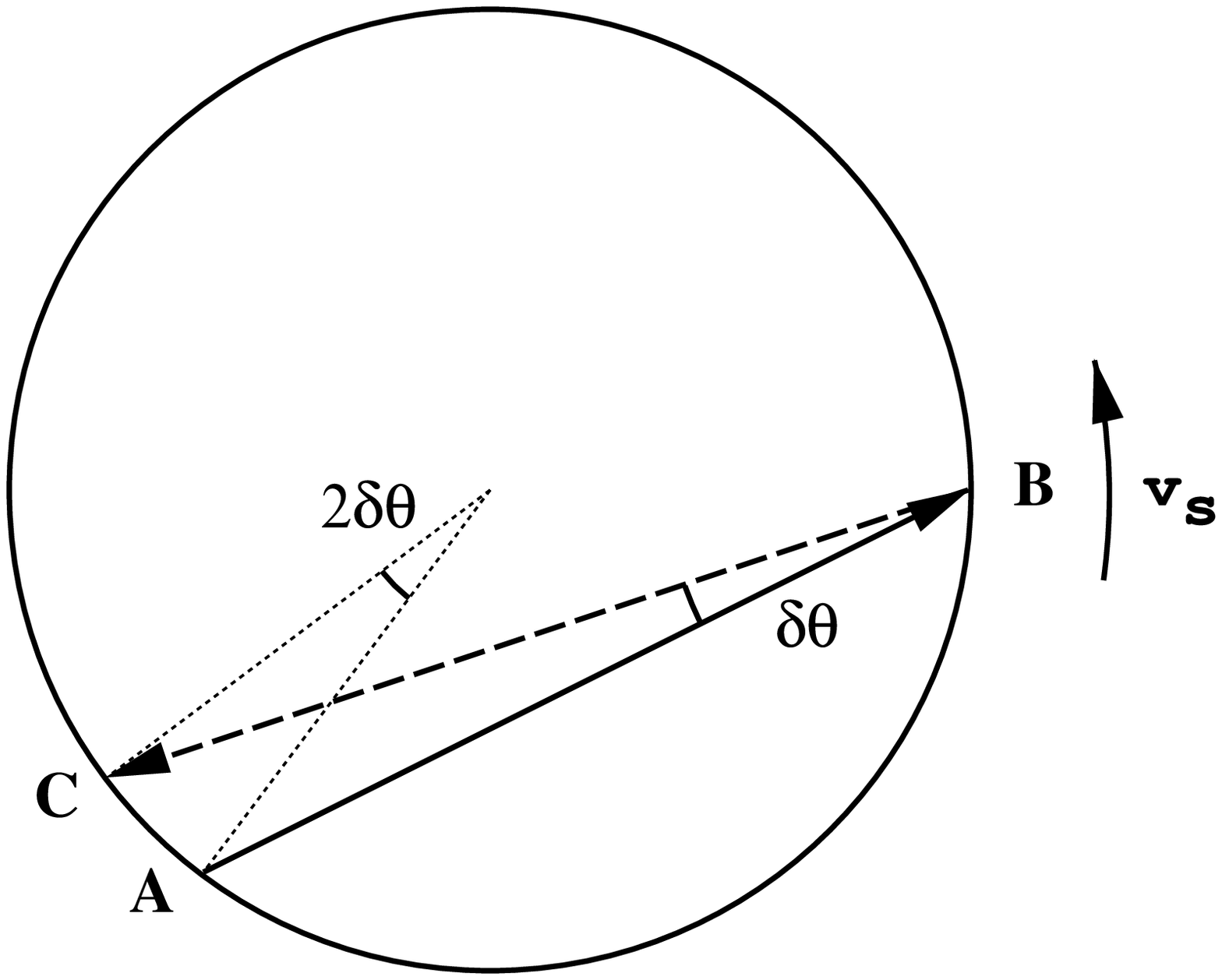}}

{\narrower\smallskip\noindent Fig. 5. Andreev reflection from the
supercurrent causes the chord defining the bound state to precess in
the opposite sense to the superflow.\smallskip}

The $l$ dependence of our core-state spectrum is therefore entirely
determined by the geometry of Andreev reflection. The only role of
the wave character of the states is in  the condition that $l$ be an
integer, and in the $n$ dependence of the spectrum.  Since only the
zero-crossing branch, $n=-1$, is of interest in the spectral flow
problem, the $n$ dependence is irrelevent here.

The lesson we learn from this section is that the quasi-classical
motion of the chord is governed by  two processes whose effects add.
One, proportional to the bound-state energy is independent of the
superflow, and the other, proportional to the superflow, is
independent of the energy. The additivity of the effects also applies
when the superflow is not parallel to the SN interface. We will need
this fact in the next section.

\line{\bf 5) Frustrated Flow and Mutual Friction\hfil}

If we insert our vortex in a transport supercurrent by replacing the
$\Delta e^{i\theta}$ gap function by $\Delta e^{i\theta}e^{2imv_s x}$
then either considerations of the group   velocity or of the geometry
of Andreev reflection will show that the
wavepackets will try to migrate across the core at speed $v_s$, just
as in the parallel sided junction.  In the absence of precession,
this motion across the core will cause a monotonic change of
$(\theta_R-\theta_L)$ from $0$ (where $\epsilon= -\Delta$) to $2\pi$
(where $\epsilon=+\Delta$) leading to a steady stream of states crossing
the gap and becoming unbound quasi-particles, just as in section 2).

This migration is, however, in competition with the $-\omega_0$ precession. 
The competition can be described quantitatively by writing an effective
hamiltonian governing the evolution of the wavepackets
$$
H= -\omega_0 l + {\bf v}_s\cdot{\bf k},
\eqno (5.1)
$$
where ${\bf k}=(k_f\cos\theta, k_f\sin\theta)$. The observables $l$
and $\theta$ are canonically conjugate and have commutation relations
$[\theta,l]=i$.

For a supercurrent in the $+x$ direction, 
the orientation and impact parameter of a wave-packet of bound states 
evolve according to the quasi-classical equations
$$
\eqalign{
\dot \theta = &\frac {\partial H}{\partial l}=-\omega_0\cr
\dot l = &-\frac {\partial H}{\partial \theta}= k_f v_s \sin \theta\cr
}
\eqno (5.2)
$$
The orbits are
$$
l=\frac{k_f v_s}{\omega_0}\cos\omega_0 t + const.
\eqno (5.3)
$$
We see that a packet with an initially positive value of 
$\sin\theta$ tries to cross the core from $-|l_{{\rm max}}|$ to $+|l_{{\rm max}}|$ 
but, because of the evolution of $\theta$, it is reflected back. This is
not surprising since the hamiltonian (5.1) is identical to the Wannier-Stark
hamiltonian for a tight-binding model. In this interpretation $l$
would label the atomic site, $\omega_0$ would be the external electric
field and $\theta$ the crystal momentum. The $2\pi$ periodicity of the latter 
is a direct consequence of the discreteness of the values of $l$ --- 
just as in the vortex the discreteness of $l$
is a direct consequence of the $2\pi$ periodicity of $\theta$.
The to-and-from motions of  $l(t)$ are therefore Bloch oscillations 
caused by the discreteness of the bound state spectrum. 

The only way a
state can make the journey from $-|l_{{\rm max}}|$ to $+|l_{{\rm max}}|$
without being reflected would be for $v_sk_f/\omega_0$ to be larger than
$l_{{\rm max}}$. But this is equivalent to the condition that $v_s$
exceed the pair-breaking velocity, $v_{\rm pair}k_f=\Delta$.

Although the spectral flow is thwarted by the Bloch oscillations
there are still effects from the evolution of the states.  We can
analyze these by writing a Boltzmann equation for the distribution
function, $n(l,\theta)$, in the $l$,$\theta$ phase space. We have
$$
\frac {\partial n}{\partial t} -\omega_0\frac {\partial n}{\partial
\theta} + k_f v_s \sin\theta\frac {\partial n}{\partial l}= I
\eqno (5.4)
$$
where $I$ is the collision integral. We will use a simple relaxation
time approximation to (5.4) by setting
$$
I= -\frac 1{\tau}(n(l,\theta)- \bar n(l,\theta))
\eqno (5.5)
$$
where $\bar n(l ,\theta)$ is an equilibrium distribution whose exact form is
determined by the normal velocity ${\bf v}_n$. (In a superconductor the
normal component relaxes to the lattice, so in this case $v_n\equiv v_{{\rm
lattice}}$.)
For example if  ${\bf v}_n= {\bf v}_s=0$ 
we would expect $n\to n_0 \equiv\theta(l)$ so that all
 negative energy states are occupied.  

It is actually more convenient to write equations for the moments of
$n$ that give the total momentum in the core
$$
\eqalign{
<k_x>=&
\frac 12 \int dl\frac {d\theta}{2\pi} (n(l,\theta)-n_0(l,\theta))
k_f\cos\theta\cr
<k_y>=&
\frac 12 \int dl\frac {d\theta}{2\pi} (n(l,\theta)-n_0(l,\theta))
k_f\sin\theta.\cr
}
\eqno (5.6)
$$
The factor of $\frac 12$ in front of these integrals 
compensates for the double counting of particles and holes, and 
$n_0(l,\theta)=\theta(l)$ reflects the
normal-ordering which ensures that $<{\bf k}>=0$ when everything is at rest.

As an example suppose we fill all the negative energy states of (5.1) by setting
$n(l,\theta)= \theta(\omega_0 l - v_s k_f \cos \theta)$. We find
$<k_y>=0$ and 
$$
<k_x>= -\frac 14 \frac {k_f^2 v_s}{\omega_0} = -\rho \kappa \frac {v_s}{\omega_0}
\eqno (5.7)
$$ 
where $\kappa= \pi/m$ is the quantum of circulation and $\rho = m k_f^2/4\pi$
is the equilibrium mass density of the electron fluid. 
This would be the appropriate equilibrium distribution 
for the case ${\bf v}_n=0$.
The non-zero value of $<k_x>$ reflects the reduction in current
through the vortex core because of the non-zero value of $\rho_n$
there. If ${\bf v}_n={\bf v}_s $, on the other hand, we expect $<{\bf k}>=0$
because, even though $\rho_n\ne 0$ in the core, the total current  
${\bf j}=\rho_s {\bf v}_s+ \rho_n {\bf v}_n$ is unaffected by the vanishing gap.

Using these definitions and insights we find
$$
\eqalign{
\frac{d<k_x>}{dt}=&\omega_0 <k_y> +\kappa\rho v_{sy}
-\frac 1\tau \left(<k_x> - <k_x>_0 \right)\cr
\frac{d<k_y>}{dt}=-&\omega_0 <k_x> -\kappa\rho v_{sx}
-\frac 1\tau \left(<k_y>-<k_y>_0 \right)\cr
}
\eqno (5.8)
$$
where
$$
<{\bf k}>_0= 
\frac {\kappa\rho}{\omega_0}({\bf v}_{n}-{\bf v}_{s}).
\eqno (5.9)
$$

In the steady state flow $<k_{x,y}>$ will be constant, so the left
hand side of (5.8) will be zero.  We can therefore solve for
$<k_{x,y}>$ and find the rate at which momentum is being transferred
to the lattice
$$
\left(\frac{d<{\bf k}>}{dt}\right)_{{\rm lattice}}=
\frac 1\tau \left(<{\bf k}>-<{\bf k}>_0 \right).
\eqno (5.10)
$$ 
We find 
$$
\eqalign{
\left(\frac{d<k_x>}{dt}\right)_{{\rm lattice}}=&-\kappa\rho \frac
{\omega_0\tau}{1+\omega_0^2\tau^2} v_{nx} +\kappa \rho \frac
{1}{1+\omega_0^2\tau^2} v_{ny}\cr
\left(\frac{d<k_y>}{dt}\right)_{{\rm lattice}}=&-\kappa\rho \frac
{\omega_0\tau}{1+\omega_0^2\tau^2} v_{ny} -\kappa \rho \frac
{1}{1+\omega_0^2\tau^2} v_{nx}.\cr
}
\eqno (5.11)
$$

\noindent Notice that ${\bf v}_s$ does not occur in (5.11).

These calculations have been made  with the vortex stationary. Because
of the galilean invariance of the overall system, we may replace ${\bf
v}_n$ by $({\bf
v}_n - {\bf v}_L)$ when the vortex is moving at velocity ${\bf v}_L$.
The rate of loss of momentum from the  core to the lattice is then
$$
\left(\frac{d<{\bf k}>}{dt}\right)_{{\rm lattice}}=  D({\bf v}_L - {\bf
v}_n) +D' \hat {\bf z} \times ({\bf v}_L - {\bf v}_n)
\eqno (5.12)
$$
The quantities  $D$, $D'$ are the mutual friction coefficients. Their
values
$$
\eqalign{
D= &\kappa \rho \frac
{\omega_0\tau}{1+\omega_0^2\tau^2} \cr
D'= &\kappa \rho \frac
{1}{1+\omega_0^2\tau^2}\cr
}
\eqno (5.13)
$$
are the same  as found, for example,  in [16].  For the
purposes of making comparisons with other work we must point out
that  in deriving these expressions  we have assumed that we are at
sufficiently low temperatures that there are very few quasi-particles
outside the vortex core. We have therefore   made no distiction
between $\rho$ and $\rho_s$. Similarly the linearization inherent in
our use of the Andreev equations makes it difficult to distinguish
between $\rho$ and the constant $C_0$ which is defined as the
electron mass times number of states within the Fermi sphere, and so
coincides with the normal-state density.

Equation (5.13) determines that Hall angle.  While the task of finding a
solution to the full dynamical gap equations for a vortex immersed in
a superflow is rather daunting, we do know that such a solution must
satisfy the law of momentum conservation.  The vortex core velocity
$v_L$
must therefore satisfy the 
momentum balance equation
$$ 
0=
\kappa\rho \hat {\bf z}\times({\bf v}_L - {\bf v}_s)  - D({\bf v}_L -
{\bf v}_n) -D' \hat {\bf z} \times ({\bf v}_L - {\bf v}_n)
\eqno (5.14) 
$$ 
as a consistency condition.  The first term is the Magnus force which
gives the rate at which momentum enters the core. The terms with $D$
and $D'$ give the rate at which momentum is being lost from the core
to the lattice by mutual friction.

There are two extreme cases.  In the collisionless  regime,
$\omega\tau \gg 1$, the vortex has no choice to move with the
superflow. In the opposite, hydrodynamic, regime $\omega_0 \tau$ is
small.  Then the ${\bf v}_L$ part of the $D'$ term almost cancels the
${\bf v}_L$ part of the Magnus force term.  This allows the vortex to
move at right-angles to the superflow, unwinding the order parameter
phase gradient and dissipating the superflow.

It is worth pointing out that the cancellation between the two $v_L$
terms in the hydrodynamic limit should {\it not} be regarded as a
``cancellation'' of the Magnus force. To claim that it is, is akin
to  saying that the the ``$F$'' cancels the ``$ma$'' in $F=ma$. This
may seem  a mere quible, but an an inappropriate choice
of language  has caused confusion in the literature.
It is also worth stressing that while the reactive term $D' \hat {\bf z}
\times ({\bf v}_L - {\bf v}_n)$ does not cause any dissipation of
energy, its presence is entirely due to quantum inchoherent,
entropy-generating relaxation processes.

\vskip 12 pt

\line{\bf 6) Conclusion and Discussion\hfil}

We have seen that a quasi-classical geometric optics model  gives a
good description of the processes in the core of a vortex immersed in
a transport current. Even though there is no explicit time dependence
in the Bogoliubov-de Gennes equations, spectral flow would still
occur were it not suppressed by an analogue of Bloch oscillations
originating in the discrete nature of the spectrum. The spectral
evolution does, however, lead to a non-equilibrium occupation of
momentum-carrying core states.  When we account for the processes by
which the occupation distribution attempts to relax, we find
expressions for  the mutual friction parameters which coincide with
those obtained by traditional Green function methods [6,7,15]. It is
therefore clear that the Green functions tacitly include the physics
of spectral flow. The new method of explicitly following the
evolution of the states does however have the advantage of being much
more physical. 

Our general conclusions are consistent with recent work by Kopnin
{\it et al.\/} which is briefly reviewed in  [17]. Some other recent
discussions of the Magnus force and momentum balance are by Hoffmann
and K{\"u}mmel [18],  Gaitan [19],  Ao [20], and {\v S}im{\'a}nek
[21]. The first of these papers has much in common with the present
approach.  In particular these authors focus on the role Andreev
scattering plays in transferring momentum from the condensate to the
core states. The next two use the Berry phase approach to confirm
that the Magnus force correctly gives the momentum flow into the
vortex core. The last combines effects of external and core states to
find an effective ``Magnus'' force that is a combination of the true
Magnus force and the mutual friction terms.

\vskip 12 pt

\line{\bf Acknowledgements\hfil}

This work was supported by the National Science Foundation  under
grant DMR94-24511.  I have benefited by conversations with Ping Ao,
Gordon Baym, Frank Gaitan, Ioan Kosztin, Tony Leggett, Qian Niu, and
Xiaomei Zhu. I must especially thank Paul Goldbart for his crucial help
in locating a missing factor of $\cos \theta$.

\vfil\eject

\line{\bf References\hfil}

\item{[1]} P.~Ao, D.~J.~Thouless, Phys.~Rev.~Lett. {\bf 70}(1993) 2158;
P.~Ao, Q.~Niu, D.~J.~Thouless, Physica {\bf B194-96} (1994) 1453.

\item{[2]} G.~K.~Batchelor, {\it An Introduction to Fluid
Dynamics\/},(Cambridge 1967)  p405.

\item{[3]} C.~Caroli, P.~G.~de Gennes, J.~Matricon, Phys.~Lett.
 {\bf 9} (1994) 307.

\item{[4]} J.~Bardeen, R.~Kummel, A.~E.~Jacobs, L.~Tewordt, 
Phys.~Rev. {\bf 187} (1969) 556

\item{[5]} J.~Bardeen, R.~D.~Sherman, Phys.~Rev. {\bf B12} 
(1975) 2634;

\item{[6]} N~B.~Kopnin, V.~E.~Kravtsov, Zh.~Eksp.~Teor.~Fiz. {\bf 
 71} (1976) 1644. [Sov. Phys. JETP 44 (1976)
861.]; N~B.~Kopnin, V.~E.~Kravtsov, Pis'ma  Zh.~Eksp.~Teor.~Fiz. {\bf
23} (1976) 631. [JETP Letters {\bf 23} (1976) 578].

\item{[7]}  A.~I.~Larkin, Yu.~N.~Ovchinnikov, 
Zh.~Eksp.~Teor.~Fiz {\bf 73}
      (1977) 299.  [Sov. Phys. JETP {\bf 46} (1977) 155.]

\item{[8]} G.~E.~ Volovik, Zh.~Eksp.~Teor.~Fiz. {\bf 104}
(1993)3070.  [Sov. Phys. JETP {\bf 77} (1993) 435.];
G.~E.~Volovik, Pis'ma  Zh.~Eksp.~Teor. Fiz. {\bf 57} (1993) 
233. [JETP Letters, {\bf 57} (1993)].

\item{[9]} Y.~G.~ Makhlin, T.~S.~Misirpashaev, Pis'ma  Zh.~Eksp.~Teor.
Fiz.{\bf 62} (1995) 74.
[JETP Letters {\bf 62} (1995) 83].

\item{[10]} A.~F.~Andreev,  Zh.~Eksp.~Teor.~Fiz.{\bf 46} (1964)
1823. [Sov. Phys. JETP {\bf 19} (1964) 1228].

\item{[11]} G.~H.~Wannier, Phys. Rev. {\bf 117} (1960) 432; For a recent account see
 M.~C.~~Chang, Q.~Niu,  Phys. Rev. {\bf B48} (1993) 2215.

\item{[12]} J.~Goldstone, F.~Wilczek, Phys.~Rev.~Lett {\bf 47} (1981)
986.

\item{[13]} M.~Stone, F.~Gaitan, Ann.~Phys. (NY) {\bf 178} (1987) 89.

\item{[14]} L.~ Kramer, W.~Pesch,  Zeitschrift fur Physik {\bf A269}
(1974) 59.

\item{[15]} S.~Hoffmann, R.~K{\"u}mmel, Z.~Phys. {\bf 84} (1991)
237.

\item{[16]} N.~B.~Kopnin, Phys.~Rev. {\bf B47} (1993) 14354.

\item{[17]} N.~B.~Kopnin, G.~E.~Volovik, Ue.~Parts,
      Europhysics Letters, {\bf 32}  (1995) 651.

\item {[18]} S.~Hoffmann, R.~K{\"u}mmel, Phys.~Rev.~Lett. {\bf 70}
1319 

\item{[19]} F.~Gaitan, Phys.~Rev. {\bf B51} (1994) 9061

\item{[20]} P.~Ao, cond-mat/9603196 (Physics Letters A, in press)

\item{[21]} E.~{\v S}im{\'a}nek. Phys.~Rev. {\bf B52} (1995) 10336
 \bye